\documentclass[preprint,12pt]{elsarticle}
\usepackage{fancyhdr}
\usepackage{algpseudocode}
\usepackage{algorithm}
\usepackage{listings}
\usepackage{tikz}
\usepackage[T1]{fontenc}
\usepackage{soul}
\usepackage{graphicx}
\usepackage{epstopdf}
\usepackage{amsmath}
\usepackage{subfigure}
\usepackage{epsfig}
\usepackage{citesort}
\usepackage{hyperref}
\usepackage{breakurl}
\usepackage{etoolbox}

\biboptions{numbers,sort&compress}

\journal{Computer Physics Communications}

\begin{document}

\begin{frontmatter}

\title{AFiD-GPU: a versatile Navier-Stokes Solver for Wall-Bounded Turbulent Flows on GPU Clusters}
\author[ut]{Xiaojue Zhu}
\author[nvd]{Everett Phillips}
\author[ut]{Vamsi Spandan}
\author[sur]{John Donners}
\author[nvd]{Gregory Ruetsch}
\author[nvd]{Josh Romero}
\author[hvd]{Rodolfo Ostilla-M\'onico}
\author[ut]{Yantao Yang} 
\author[ut,mkp]{Detlef Lohse}
\author[rom,ut]{Roberto Verzicco}
\author[nvd]{Massimiliano Fatica}
\author[ut]{Richard J.A.M. Stevens}

\address[ut]{Physics of Fluids Group, MESA+ Institute, and J. M. Burgers Centre for Fluid Dynamics,\\ University of Twente, PO Box 217, 7500 AE Enschede, The Netherlands.}
\address[nvd]{NVIDIA Corporation, 2701 San Tomas Expressway,Santa Clara, CA 95050, USA}
\address[sur]{SURFsara, Science Park 140, 1098 XG Amsterdam, The Netherlands}
\address[hvd]{School of Engineering and Applied Sciences and Kavli Institute for Bionano Science and Technology,\\\ Harvard University, Cambridge, MA 02138, USA}
\address[rom]{Dipartimento di Ingegneria Industriale, University of Rome ``Tor Vergata'', Via del Politecnico 1, Roma 00133, Italy}
\address[mkp]{Max-Planck Institute for Dynamics and Self-Organization, Am Fassberg 17, 37077 G\"{o}ttingen, Germany. }

\begin{abstract}
The AFiD code, an open source solver for the incompressible Navier-Stokes equations ({\color{blue}\burl{http://www.afid.eu}}), has been ported to GPU clusters to tackle large-scale wall-bounded turbulent flow simulations. The GPU porting has been carried out in CUDA Fortran with the extensive use of kernel loop directives (CUF kernels) in order to have a source code as close as possible to the original CPU version; just a few routines have been manually rewritten. A new transpose scheme, which is not limited to the GPU version only and can be generally applied to any CFD code that uses pencil distributed parallelization, has been devised to improve the scaling of the Poisson solver, the main bottleneck of incompressible solvers. The GPU version can reduce the wall clock time by an order of magnitude compared to the CPU version for large meshes. Due to the increased performance and efficient use of memory, the GPU version of AFiD can perform simulations in parameter ranges that are unprecedented in thermally-driven wall-bounded turbulence. To verify the accuracy of the code, turbulent Rayleigh-B\'enard convection and plane Couette flow are simulated and the results are in good agreement with the experimental and computational data that published in previous literatures. \end{abstract}


\end{frontmatter}

{\bf PROGRAM SUMMARY}

\begin{small}
\noindent
{\em Program Title:}         AFiD-GPU                                 \\
{\em Licensing provisions(please choose one): GPLv3 }                                   \\
{\em Programming language:} Fortan 90, CUDA Fortan, MPI                                  \\
{\em External routines:}  PGI, CUDA Toolkit, FFTW3, HDF5                               \\
{\em Nature of problem(approx. 50-250 words):} Solving the three-dimensional Navier-Stokes equations coupled
 with a scalar field in a cubic box bounded between two walls and other four periodic boundaries.\\
{\em Solution method(approx. 50-250 words):} Second order finite difference method for spatial discretization, third order Runge-Kutta scheme and Crank-Nicolson method for time advancement, two dimensional pencil distributed MPI parallelization, GPU accelerated routines.\\
{\em Additional comments including Restrictions and Unusual features (approx. 50-250 words):} The open-source code is supported and updated on {\color{blue}\burl{http://www.afid.eu}}.\\
   \\

\end{small}

\section{Introduction}

Turbulence is a high dimensional multi-scale process. As the velocity of the fluid increases, the range of scales of the resulting motion increases as energy is transferred to smaller and smaller scales, and the flow transitions from laminar to turbulence. To understand the physics of this energy transfer, Direct Numerical Simulations (DNS) are used that resolve all of these scales. In order to resolve scales, large meshes and immense computational power are required. 

Here, two paradigmatic systems are taken as examples, i.e. Rayleigh-B\'enard convection \cite{ahl09,loh10,chi12}, the buoyancy driven flow of a fluid heated from below and cooled from above, and plane Couette flow, the shear-induced motion of a fluid contained between two infinite flat walls, which are among the most popular systems for convection and wall-bounded shear flow. The two are classical problems in fluid dynamics. Next to pipe \cite{eck07}, channel \cite{kim87,lee15}, and Taylor-Couette flows \cite{far14,gro16}, the systems have been and are still used to test various new concepts in the field \cite{ahl09} such as nonlinear dynamics and chaos, pattern formation, or turbulence, on which we focus here. 

Turbulent Rayleigh-B\'enard flow is of interest in a wide range of sciences, including geology, oceanography, climatology, and astrophysics as it is a relevant model for countless phenomena such as thermal convection in the atmosphere \cite{har01}, in the oceans (including thermohaline convection) \cite{mar99}, in Earth's outer core \cite{car94}, where the reversals of the large scale convection are of prime importance to the magnetic field, in the interior of gaseous giant planets and in the outer layer of the Sun \cite{cat03}. Natural convection in technological applications such as buildings, in process technology, or in metal-production processes is also modeled using Rayleigh-B\'enard flow. For those real-world applications of Rayleigh-B\'enard flow, the system is highly turbulent in both bulk and boundary layers. This state is the so-called ultimate regime of thermal convection, which has been recently realized experimentally in the laboratory  \cite{he12}. However, because of the extremely high Rayleigh numbers (the non-dimensional temperature difference) and high Reynolds numbers (the non-dimensional velocity) of the flow, computationally the ultimate regime could not be reached so far, despite its great importance.

Turbulent plane Couette flow is of interest for more fundamental reasons. It is the only flow which bears exactly the same total stress across the thickness, and which is one of hypotheses requested by Prandtl's classical arguments for the existence of logarithmic layer for the mean velocity profile \cite{pir14}. Besides, because its simple geometry, plane Coutte flow is often used as an example to illustrate the wall-bounded turbulence structure \cite{sch00}, and more recently, investigation of the self-sustainment of near wall turbulence \cite{wal97b} or inner-outer wall turbulence interaction \cite{pir11}.

To accurately simulate high Rayleigh and high Reynolds number flows of interest in geo- and astrophysical flows \cite{mck74,car94,cat03}, efficient code parallelization and effective use of large scale supercomputers are essential to reach the amount of grid points necessary to resolve all flow scales. Previous work in parallelizing a second-order finite-difference solver for natural convection and shear flow have allowed us to consider unprecedented large computational boxes using AFiD \cite{poe15cf,ste17}. However, there are still limitations to the parallelization as it was written for a central processing unit (CPU)-based system, while the current trends in High Performance Computing points towards the increase in use of accelerators. These are expected to push the performance of supercomputers into the ExaScale range by the use of graphic processing units (GPUs) \cite{lee2010}. GPUs are especially well-suited to address problems that can be expressed as data-parallel computations, where the same program is executed on different data elements in parallel. GPUs are also characterized by high memory bandwidth, something especially important for low-order finite difference computational fluid dynamics (CFD) codes where the data reuse is minimal.  Given the above and that GPUs are the most used accelerator technology, we decided to port AFiD to GPU clusters, while further developing the underlying algorithms. With the porting of AFiD to GPU, and the introduced efficiency improvements, this open source code can now tackle unprecedentedly large fluid dynamics simulations, and thus is expected to be of benefit to the convection and scientific community at large.

In \textsection \ref{AFiDcode} we discuss the details of the solver AFiD. Subsequently, in \textsection \ref{GPUimplementation} details about the GPU implementation are discussed, before we discuss the code performance in \textsection \ref{section_performance}. In \textsection \ref{section_validation} we end with a presentation of Rayleigh-B\'enard and plane Couette cases that have been simulated with the new GPU code. In section \textsection \ref{section_conclusions} we present the main conclusions and present future development plans for the code.

\section{AFiD code} \label{AFiDcode}
Here we summarize the numerical method (\textsection \ref{AFiD_method}) and the parallelization scheme (\textsection \ref{AFid_Parallelization}) as described in Ref.\ \cite{poe15cf} before we will discuss the specifics of the GPU implementation in \textsection \ref{GPUimplementation}.

\subsection{Numerical scheme} \label{AFiD_method}
AFiD ({\color{blue}\burl{http://www.afid.eu}}) solves the Navier-Stokes equations with an additional equation for temperature in three-dimensional coordinates on a Cartesian mesh with two periodic (unbounded) directions ($y$ and $z$) which are uniformly discretized and one bounded direction ($x$) for which non-uniform grids, with clustering of points near the walls, can be used. Note that for Rayleigh-B\'enard flow, the temperature is turned on and for plane Couette flow, the temperature equation is turned off and all the other features, except the boundary conditions, are the same. 

The Navier-Stokes equations with the incompressibility condition read:
\begin{equation}
\nabla \cdot \textbf{u} = 0,
\label{eq:Ns1}
\end{equation}
\begin{equation}
\displaystyle\frac{\partial\textbf{u}}{\partial t} + \textbf{u} \cdot \nabla \textbf{u} = -\rho^{-1} \nabla p 
 + \nu \nabla^2\textbf{u} + \textbf{F}_b.
\label{eq:Ns2}
\end{equation}

\noindent for the temperature field, an advection-diffusion equation is used

\begin{equation}
\displaystyle\frac{\partial T}{\partial t} + \textbf{u}\cdot \nabla T = \kappa \nabla^2 T,
\label{eq:Ns3}
\end{equation}
where $\textbf{u}$ is the velocity vector, $p$ the pressure, $\rho$ the density, $T$ the temperature, $\nu$ the kinematic viscosity, $\kappa$ the thermal diffusivity and $t$ time. For Rayleigh-B\'enard convection we use the Boussinesq approximation: the body force $F_b$ is taken to only depend linearly on the temperature, and to be in the direction of gravity ($\textbf{e}_x$ is the unit vector anti--parallel to gravity), and we also ignore the possible dependencies of density, viscosity and thermal diffusivity on temperature. Other body forces, like the Coriolis force, can be included in this term if one is dealing with rotating frames. 

\begin{figure}[!tb]
\centering
\hspace{1cm}
\begin{tikzpicture}
\draw (0,0) rectangle (4,4);
\filldraw[black] (2,2) circle(1mm);
\node [above right,black] at (2,2) {$p$};
\filldraw[black] (0,2) circle(1mm);
\node [above right,black] at (0,2) {$u_y$};
\filldraw[black] (4,2) circle(1mm);
\node [above right,black] at (4,2) {$u_y$};
\filldraw[black] (2,0) circle(1mm);
\node [above right,black] at (2,0) {$u_x,T$};
\filldraw[black] (2,4) circle(1mm);
\node [above right,black] at (2,4) {$u_x,T$};
\draw [->,thick] (4.5,0) -- (5.5,0);
\draw [->,thick] (4.5,0) -- (4.5,1.0);
\draw [->,thin, dashed] (4.5,0) -- (5.2,0.7);
\node [above,black] at (4.5,1.0) {$x$};
\node [right,black] at (5.5,0) {$y$};
\node [above right,black] at (5.2,0.7) {$z$};
\end{tikzpicture}
\caption{Location of pressure, temperature and velocities of a 2D simulation cell. The third dimension ($z$) is omitted for clarity. As on an ordinary staggered scheme, the velocity vectors are placed on the borders of the cell and pressure is placed in the cell center. The temperature is placed at the same location as the vertical velocity, to ensure exact energy conservation.}
\label{fig:disc}
\end{figure}
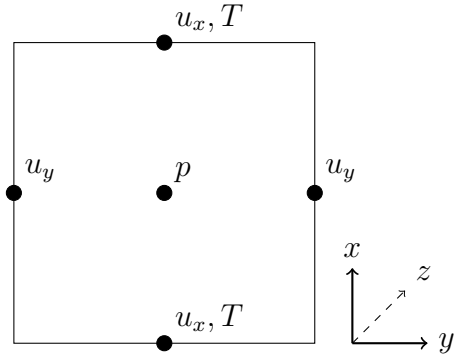

For the spatial discretization of the domain, we use a conservative, central, second--order, finite--difference discretization on a staggered grid. A two-dimensional (for clarity) schematic of the variable arrangement is shown in Fig.~\ref{fig:disc}. The pressure is calculated at the center of the cell. For thermal convection between two plates, the temperature field is collocated with the $u_x$ grid, the velocity component in the direction of gravity. This avoids the interpolation error when calculating the term $\textbf{F}_b \sim T\textbf{e}_x$ in Eq.\ \ref{eq:Ns2}. This scheme has the advantage of being energy conserving in the limit $\Delta t\to0$ \cite{orl12}. In addition to the conservation properties, the low-order finite difference scheme has the advantage of handling better the shock-like behavior resulting from the absence of the pressure term in the temperature equation from the Boussinesq approximation \cite{ost15,moi16}.

Given a set of initial conditions, the simulations are advanced in time by a fractional--step procedure combined with a low--storage, third--order Runge--Kutta (RK3) scheme and a Crank--Nicolson method \cite{rai91} for the implicit terms. The theoretical stability limit of the RK3 methods is a Courant-Friedrichs-Lewy (CFL) number equal to $\sqrt{3}$ even if, in practice, simulations are run at a maximum CFL number of approximately $1.3$. The RK3 scheme requires three substeps per time-step, but due to the larger time-step and the $\mathcal{O}([\Delta t]^3)$ error it is more efficient than a standard second--order Adams--Bashforth integration. The pressure gradient is introduced through the ``delta'' form of the pressure \cite{lee01}: a provisional, non--solenoidal velocity field is calculated using the old value of the pressure in the discretized Navier-Stokes equation. The updated pressure, required to enforce the continuity equation at every cell, is then computed by solving a Poisson equation for the pressure correction. The velocity and pressure fields are then updated using this correction, which results in a divergence--free velocity field. Full details of the procedure can be found in Ref.\ \cite{ver96}.

\subsection{Parallelization strategy} \label{AFid_Parallelization}
\begin{figure}[!tb]
 \centering
 \includegraphics[width=0.7\textwidth]{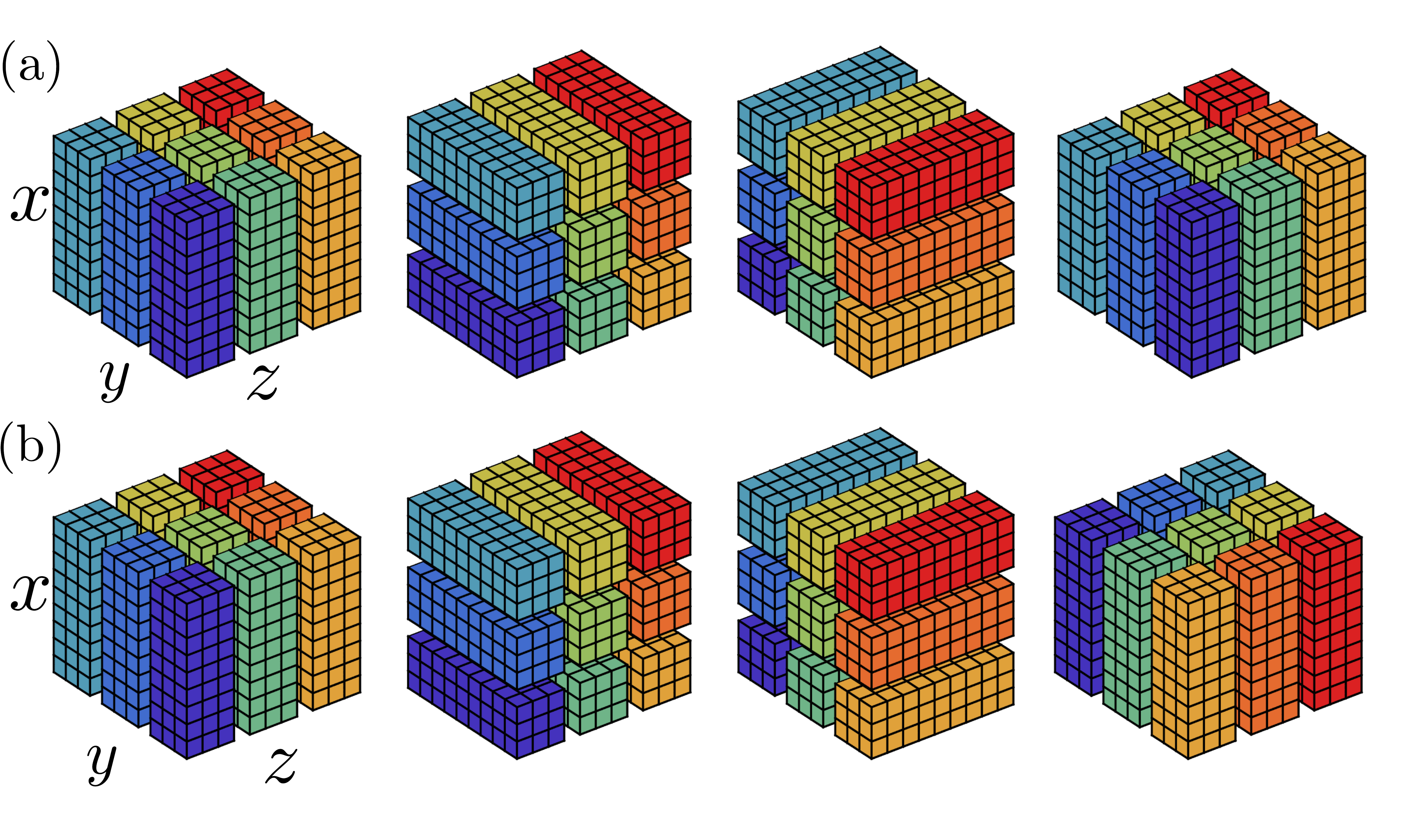}
\caption{Comparison between the original transpose scheme implemented in the CPU version (top, see \textsection \ref{AFid_Parallelization}) and the one implemented in the GPU version (bottom, see \textsection \ref{Transpose}). Both transpose strategies start from the left-most configuration, operate on the arrays, and transpose again ending up at the right-most configuration, solve the Poisson equations, and transpose data back to the left-most configuration. }
\label{fig:transGPU}
\end{figure}

The 2DECOMP \cite{li10} library is used to implement a two-dimensional domain decomposition, also known as ``pencil'' decomposition. We have extended the 2DECOMP library to suit the specifics of our scheme. For a pencil decomposition solving tridiagonal matrices in directions which the pencils are not oriented in the direction of differentiation, requires re-orienting the pencils, and thus large all-to-all communications. Ref.\ \cite{poe15cf} showed that for highly turbulent high Rayleigh number flows, using the CFL time--step constraint is sufficient to assure stability as the non-linear CFL constraint in the time--marching algorithm, inherently satisfies the stability constraints imposed by the explicit integration of the horizontal components of the viscous terms. We can thus avoid the solution of the tridiagonal matrices in the horizontal directions by integrating not only the advection terms but also the horizontal viscous terms explicitly. This makes the calculation local in space for two horizontal directions, and all-to-all communications are avoided by aligning the pencils in the wall-normal ($x$) direction. In this way, every processor possesses data from $x_1$ to $x_N$ (cf. Fig.\ \ref{fig:transGPU}) and, for every pair ($y$,$z$), a single processor has the full $x$ information needed to solve the implicit equation in $x$ without further communication. We note that halo updates must still be performed during the computation of the intermediate velocity, but this memory distribution completely eliminates the all-to-all communications.

All-to-all communications are unavoidable during the pressure correction step, as a Poisson equation must be solved. Since the two wall-parallel directions are homogeneous and periodic, it is natural to solve the Poisson equation using a Fourier expansion in two dimensions. Modified wavenumbers are used, instead of the real ones, to prevent the Laplacian from having higher accuracy in some directions~\cite{moi16}. In the limit of infinite points, i.e.~$\Delta y\to 0$, the modified wavenumbers converge to the real wavenumbers. In the CPU version, the Fast Fourier Transforms are performed using the open source FFTW~({\color{blue}\burl{http://www.fftw.org/}}) library.

By using a second--order approximation for the partial derivatives in the wall-bounded directions, the Poisson equation is reduced by a two--dimensional fast Fourier transforms to a series of one--dimensional Poisson equations that are easily inverted by a tridiagonal Thomas solver. This allows for the direct
solution of the Poisson equation in a single step, with a residual round--off-error velocity divergence ($\mathcal{O}(10^{-13})$ in double--precision arithmetics) within $\mathcal{O}(N_x N_y N_z \log[N_y] \log[N_z])$ time complexity. Due to the domain decomposition, several data transposes must be performed during the computation of the equation. The algorithm for solving the Poisson equation is as follows: 

\begin{enumerate}
 \item Calculate the local divergence from the $x$-decomposed velocities.
 \item Transpose the result of {\it 1)} from a $x$-decomposition to a $y$-decomposition.
 \item Perform a real-to-complex Fourier transform on {\it 2)} in the $y$-direction.
 \item Transpose {\it 3)} from a $y$-decomposition to $z$-decomposition.
 \item Perform a complex-to-complex Fourier transform on {\it 4)} in the $z$-direction.
 \item Transpose {\it 5)} from a $z$-decomposition to a $x$-decomposition.
 \item Solve the linear system with a tridiagonal solver in the $x$-direction.
 \item Transpose the result of {\it 7)} from a $x$-decomposition to a $z$-decomposition.
 \item Perform a complex-to-complex inverse Fourier transform on {\it 8)} in $z$-direction.
 \item Transpose {\it 9)} from a $z$-decomposition to a $y$-decomposition.
 \item Perform a complex-to-real inverse Fourier transform on {\it 10)} in a $y$ direction.
 \item Transpose {\it 11)} from a $y$-decomposition to a $x$-decomposition.
\end{enumerate}

The last step outputs the scalar correction $\phi$ in real space, decomposed in $x$-oriented pencils. Therefore, once the Poisson equation is solved, the corrected velocities and pressures are computed directly. The temperature and other scalars are advected and the time sub-step is completed. For full details of the equations solved, we refer the reader to Ref.\ \cite{poe15cf}. The algorithm outlined above only transposes one 3D array, making it very efficient. The top row of figure \ref{fig:transGPU} shows a schematic of the data arrangement and the transposes needed to implement the algorithm in the original CPU code. We wish to highlight that this algorithm uses all possible combinations of data transposes. It can be seen from Fig.\ \ref{fig:transGPU} that the $x$ to $z$ transposes and the $z$ to $x$ transposes need a more complex structure, as a process may need to transfer data to other processes which are not immediate neighbors. These transposes are absent in the 2DECOMP library on which we build. These transposes have been implemented in the CPU version using the more flexible all-to-all calls of the type {\tt MPI$\_$Neighbor$\_$alltoallw} available in {\tt MPI 3.0}, instead of the all-to-all MPI calls of the type {\tt ALLTOALLV} used for the other four transposes. The GPU version uses a different transpose scheme that will be described later.

\section{GPU implementation} \label{GPUimplementation}
In this section we explain the details of the GPU implementation of AFiD. It is now possible to program GPUs in several languages, from the original CUDA C to the new OpenACC directive based compilers. We decided to use CUDA Fortran (\textsection \ref{CUDAFortran}) as the nature of the code, where most routines are nested do-loops, allows the extensive use of CUF kernels (\textsection \ref{CUFkernels}, kernel loop directives), making the effort comparable to an OpenACC port, while also retaining the possibility of using explicit code kernels when needed. In addition, the explicit nature of data movement in CUDA Fortran allows us to better optimize the CPU/GPU data movement and network traffic, and to further increase code performance. In section \textsection \ref{Memory} we describe the optimization of memory usage and the new improved transpose scheme is introduced in \textsection \ref{Transpose}. The multi-GPU aspects (\textsection \ref{MultiGPU}) and the code profiling methods (\textsection \ref{Profiling}) that have been used. 

\subsection{CUDA \& CUDA Fortran} \label{CUDAFortran}

CUDA-enabled GPUs can contain anything from a few to thousands of processor cores which are capable of running tens of thousands of threads concurrently. To allow for the same CUDA code to run efficiently on different GPUs with varying number of resources, a hierarchy of resources exists both in physical hardware, and in available programming models. In hardware, the processor cores on a GPU are grouped into multiprocessors. The programming model mimics this grouping: a subroutine, called a kernel, which runs on the device, is launched with a grid of threads grouped into thread blocks. Within a thread block data can be shared between threads, and there is a fine-grained thread and data parallelism. Thread blocks run independently of one another, which allows for scalability in the programming model: each block of threads can be scheduled on any of the available multiprocessors within a GPU, in any order, concurrently or sequentially, so that a compiled CUDA program can execute on a device with any number of multiprocessors. This scheduling is performed behind the scenes, the CUDA programmer needs only to partition the problem into coarse sub-problems that can be solved independently in parallel by blocks of threads, where each sub-problem is solved cooperatively in parallel by all threads within the block. The CUDA platform enables hybrid computing, where both the host (CPU and its memory) and device (GPU and its memory) can be used to perform computations. From a performance perspective, the bandwidth of the PCI bus is over an order of magnitude less than the bandwidth between the device's memory and GPU, and therefore a special emphasis needs to be placed on limiting and hiding PCI traffic. For MPI applications, data transfers between the host and device are required to transfer data between MPI processes. Therefore, the use of asynchronous data transfers, i.e.\ performing data transfers concurrently with computations, becomes mandatory. 

CUDA Fortran is essentially regular Fortran with a handful of extensions that allow portions of the computation to be off-loaded to the GPU. There are two compilers, at the moment, that are able to parse these extensions, the PGI compiler (now freely available via the community edition) and the IBM XLF compiler, which currently implements a subset of CUDA Fortran, in particular it does not have CUF kernels. Because we rely heavily on CUF kernels in our GPU implementation, all the results presented in this paper are obtained with the PGI compiler. CUDA Fortran has a series of extensions, like the variable attribute {\tt device} used when declaring data that resides in GPU memory, the new F2003 {\tt sourced} allocation construct and the flexibility of kernels which make porting much easier.

CUDA Fortran can automatically generate and invoke kernel code from a region of host code containing tightly nested loops. Such code is referred to as a CUF kernel. One can port code to the device using CUF kernels without modifying the contents of the loops using the following programming convention. The directive will appear as a comment to the compiler if GPU code generation is disabled or if the compiler does not support them (similar to the OpenMP directives that are ignored if OpenMP is not enabled). The contents of the loop are usually unaltered.

\subsection{CUF kernels} \label{CUFkernels}

One of the project goals was to have a code as close as possible to the original CPU version. In order to accomplish this the GPU implementation makes extensive use of the preprocessor and all the GPU specific code and directives are guarded by {\em USE\_CUDA} macro. For the same F90 source file, a CPU object file can be created with the standard optimization flags while a GPU version can be created adding the "-O3 -DUSE\_CUDA -Mcuda" flags. While the GPU code needs to be compiled with the PGI compiler, the CPU code can be compiled with any Fortran compiler. The build system will build a copy of the code for GPU and one for CPU. The original CPU code uses custom allocators that allocate and initialize to zero the arrays. Some arrays are defined with halo cells, others only for the interior points. The 2DECOMP \cite{li10} library is also using global starting indices. In order to make identical copies on the GPU, we used the F2003 {\em sourced} allocation construct. 

\begin{center}
\begin{minipage}{.75\textwidth}
 \lstset{frame=single,
 tabsize=2,
 numbers=none, 
 numbersep=5pt, 
 captionpos=b,
 basicstyle=\footnotesize}
\begin{lstlisting}[caption=Routine to compute the maximum CFL number.]
 subroutine CalcMaxCFL(cflm)
#ifdef USE_CUDA
 use cudafor
 use param,only:fp_kind,nxm, &
                dy=>dy_d,dz=>dz_d, &
                udx3m=>udx3m_d
 use local_arrays,only:vx=>vx_d, &
                       vy=>vy_d, &
                       vz=>vz_d
#else
 use param,only:fp_kind,nxm,dy,dz,udx3m
 use local_arrays,only:vx,vy,vz
#endif
 use decomp_2d
 use mpih
 implicit none
 real(fp_kind),intent(out):: cflm
 integer::i,j,k,ip,jp,kp
 real(fp_kind)::qcf

 cflm=real(0.00000001,fp_kind)

!$cuf kernel do(3) <<<*,*>>>
 do i=xstart(3),xend(3)
  ip=i+1
  do j=xstart(2),xend(2)
   jp=j+1
   do k=1,nxm
    kp=k+1
    qcf=( abs((vz(k,j,i)+vz(k,j,ip))*dz) &
         +abs((vy(k,j,i)+vy(k,jp,i))*dy) &
         +abs((vx(k,j,i)+vx(kp,j,i))*udx3m(k)))
          
    cflm = max(cflm,qcf*0.5_fp_kind)
   enddo
  enddo
 enddo

 call MpiAllMaxRealScalar(cflm)
 return  
 end
\end{lstlisting}
\end{minipage}
\end{center}

Listing 1 shows an example of the changes to the original source code enabled when -DUSE\_CUDA is passed to the compiler. As we can see from the source code, the CUF kernel directives are very simple to use. Once the compiler is aware that the 3 nested do loops need to be parallelized, it automatically determines that {\tt cflm} requires a reduction. Using the renaming facilities when loading the variables from the module, we ensure that the CUF kernel will operate on arrays resident in GPU memory. We used CUF kernels extensively, and only some routines are coded manually on the GPU. One of these routines is the routine computing the statistics, since the reduction operator is on a vector (the statistics are accumulated only in the direction normal to the wall) and, at the moment, CUF reductions only work on scalars. We also wrote batched tridiagonal solvers, where each thread in a block solves a system, and routines for local transposes required to optimize the memory layout before tridiagonal solvers or FFT (for which we used the CUFFT library).

\subsection{Reducing the memory footprint} \label{Memory}

With the computational power of the GPUs, the memory footprint becomes the limiting factor in increasing the resolution of the simulations. Reducing the memory footprint becomes one of the main objectives. While there are now GPUs with up to 24GB of memory, in most large systems the GPUs are older, and have less capacity, being 6GB and 12GB the typical size. Thus, even reducing the number of 3D arrays by a single unit results in relevant benefits. Since several routines require storing data that is only needed temporarily, either as an intermediate result or to transform the data layout, we are able to reduce the memory footprint by reusing the memory for these arrays as much as possible. In particular, there are two work arrays used in the Poisson solver that are used for either complex or real data types. In Fortran77, it was possible to use the {\tt equivalence} statement to have the two arrays sharing the same memory. While {\tt equivalence} is still supported (but deprecated) in Fortran90, it only works with statically defined arrays and the memory allocation in AFiD is all dynamic. Using the {\tt iso\_c\_binding}, it is possible to reproduce the behavior of {\tt equivalence}:

\small
\begin{verbatim}
complex, target, allocatable:: complex_vec(:)
type(c_ptr):: cptr
real, pointer:: real_vec(:)

allocate(complex_vec(N))
cptr=c_loc(complex_vec)
call c_f_pointer(cptr,real_vec, &
 [2*size(complex_vec,1)])
\end{verbatim}
\normalsize

This approach works for both CPU and GPU arrays (if the arrays are declared with the {\tt device} attribute). Another area where memory can be reduced is in the workspace that is used by the FFT library. When creating an FFT plan with CUFFT, a workspace is allocated by the library which is roughly the same size as the data that will be processed by the plan. Since the four FFT plans in the solver will not be used simultaneously, we can reuse the same storage for all the workspaces by creating the FFT plans with the new CUFFT plan management API that allows the programmer to provide the workspace memory. The initial GPU version of the code needed 48 K20x to run a $1024^3$ grid, the final version can now run on 25 K20x with 6GB of memory.

\subsection{Multi GPU implementation} \label{MultiGPU}
In the GPU implementation, we map each MPI rank to a GPU. The code discovers the available GPUs on each node and makes a 1:1 mapping between ranks and GPUs, as described in \cite{CUDAFortranBook}. In the basic version of the code the whole computation is performed on the GPUs, the CPUs are only used for I/O and to stage the data needed during the communication phases. There are MPI implementations that are GPU-aware and allow to use data resident in GPU memory directly in MPI calls, but for this initial version we used standard MPI to have a more portable code, so the data needs to be resident in CPU memory before the MPI calls. Instead of using {\tt MPI\_ALLTOALL} or {\tt MPI\_NEIGHBOR\_ALLTOALLW} calls, we used a combination of {\tt IRECV/ISEND} together with {\tt cudaMemcpy2DAsync} to better overlap transfer to/from GPU memory from/to CPU memory and communications \cite{MultiGPU}.

\begin{figure*}
\begin{centering}
\includegraphics[width=1.\textwidth]{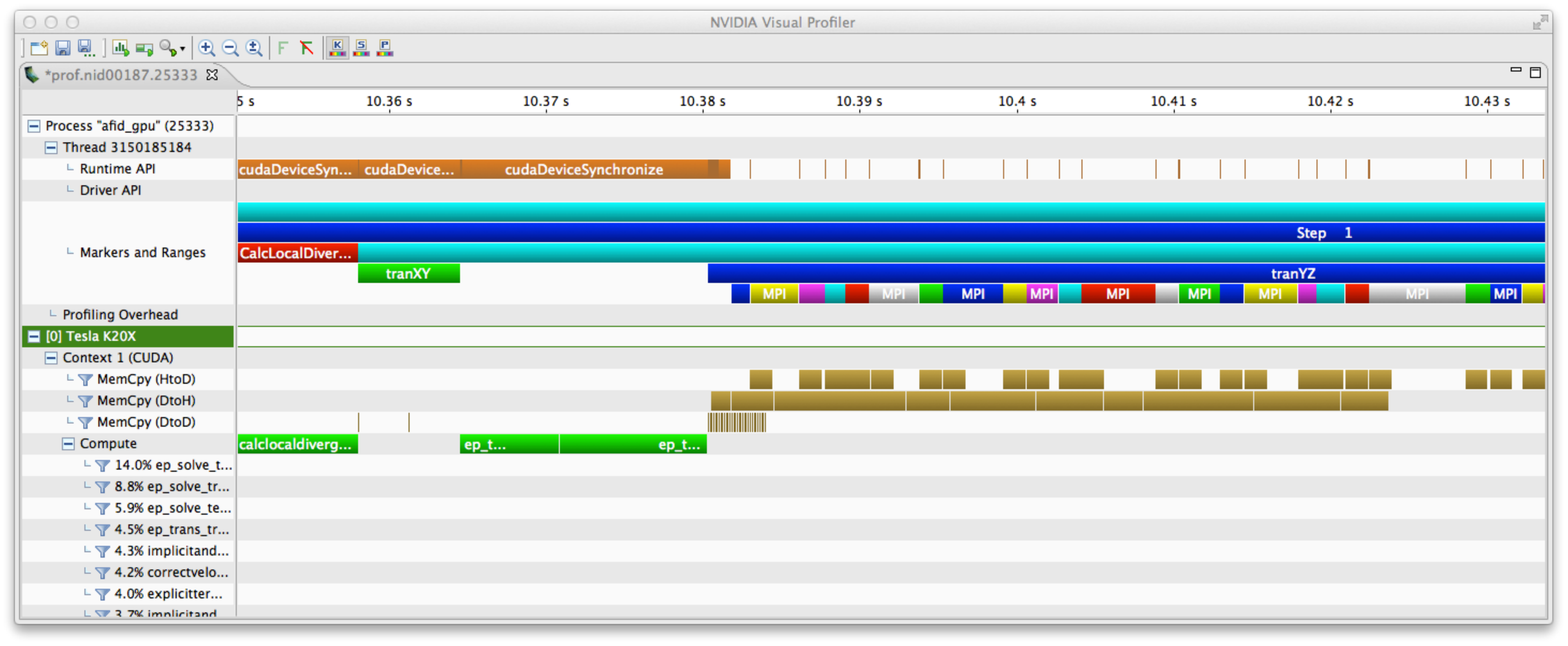}
\caption{Profiler output from AFiD\_gpu for a parallel run on a $1024^3$ grid.} \label{fig:nvvp2} 
\end{centering}
\end{figure*}

\subsection{Efficient data transposes} \label{Transpose}
AFiD was designed for high Reynolds number simulations and its parallel implementation is deeply tied to the underlying numerical scheme. As explained in \textsection \ref{AFid_Parallelization}, the code only needs to solve implicitly in the wall normal direction. AFiD uses a two-dimensional pencil decomposition aligned in the wall normal direction. Per time-step, only six all-to-all communications are required, and these are all found in the Poisson solver for the pressure correction. The original 2DECOMP library had only four transposes available (no x-to-z and z-to-x, since the library was designed for full spectral solvers for which there is no need to go back to the original vertical decomposition). The AFiD code added the x-to-z and z-to-x transposes using the new {\tt MPI 3.0 MPI$\_$Neighbor$\_$alltoallw} calls. Since in the GPU implementation we want to use combination of {\tt IRECV/ISEND} calls that allow a better overlap of data transfer from/to GPU, this required a new transpose scheme. If we relax the constraint that the tridiagonal solvers are solved in a decomposition identical to the original one in which the right hand side (local divergence) of the Poisson equation was computed, we can devise a more efficient transpose. As shown in the bottom part of Fig.\ \ref{fig:transGPU}, if we apply another rotation from $z$ to $x$ (similar to what we would do with a Rubik's cube), each processor will only exchange data with other processors in the same row sub-communicator, similar to the previous stages and use combination of {\tt IRECV/ISEND} calls. We are still using the 2DECOMP library to do the book keeping, and since the library uses global indices for addressing, we just need to access the proper wave numbers to solve the tridiagonal systems. 

\subsection{Profiling using NVTX} \label{Profiling}
Profiling is an essential part of performance tuning used to identify parts of the code that may require additional attention. When dealing with GPU codes, profiling is even more important as new opportunities for better interactions between the CPUs and the GPUs can be discovered. The standard profiling tools in CUDA, nvprof and nvvp, are able to show the GPU timeline but do not present CPU activity. The NVIDIA Tools Extension (NVTX) is a C-based API (application program interface) to annotate the profiler time line with events and ranges and to customize their appearance and assign names to resources such as CPU threads and devices \cite{NVTX}.

We have written a Fortran module to instrument CUDA/OpenACC Fortran codes using the Fortran ISO C bindings \cite{NVTXFortran}. To eliminate profiling overhead during production runs, we use a preprocessor variable to make the profiling calls return immediately. During the runs, one or more MPI processes generate the traces that are later imported and visualized with nvvp, the NVIDIA Visual Profiler. Fig.\ \ref{fig:nvvp2} shows an example of the output for AFiD\_GPU on a $1024^3$ mesh, where on the top part ``process AFiD GPU" the CPU sections can be identified while the GPU sections are on the lower ``Tesla K20x" section. The profiler is visualizing the output from one of the ranks. Since the run was on a $1 \times 16$ processor grid, we can see that after the computation of the local divergence (red box labeled CalcLocal) the first transpose, TranXY, does not require MPI communications. The following one, TranYZ, requires MPI communications and we can see the overlapping of Memcopy DtoH (device to host) and HtoD (host to device) with MPI calls. 

\section{Code performance } \label{section_performance}
In this section we first explain in \textsection \ref{OptimalConfiguration} how the two dimensional decomposition is used before we explain the detailed performance tests in \textsection \ref{PerformanceComparison}. 

\subsection{Optimal configuration} \label{OptimalConfiguration}

\begin{figure}[!tb]
\begin{centering}
\includegraphics[width=0.65\textwidth]{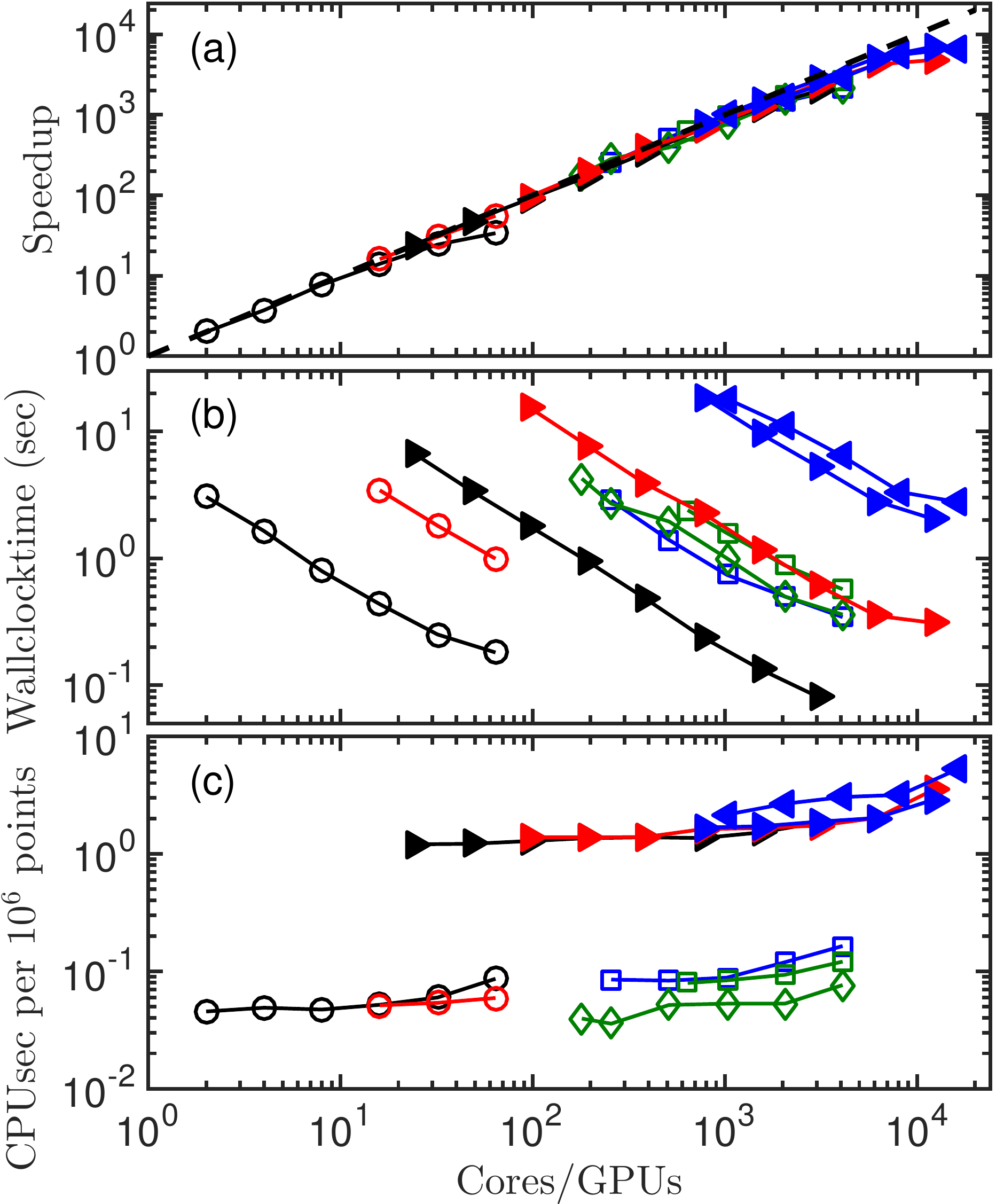}
\includegraphics[width=0.37\textwidth]{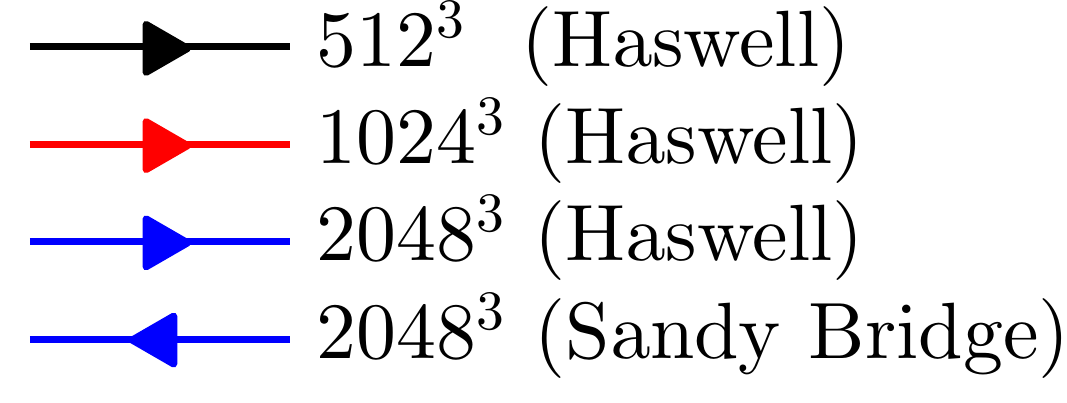}
\includegraphics[width=0.37\textwidth]{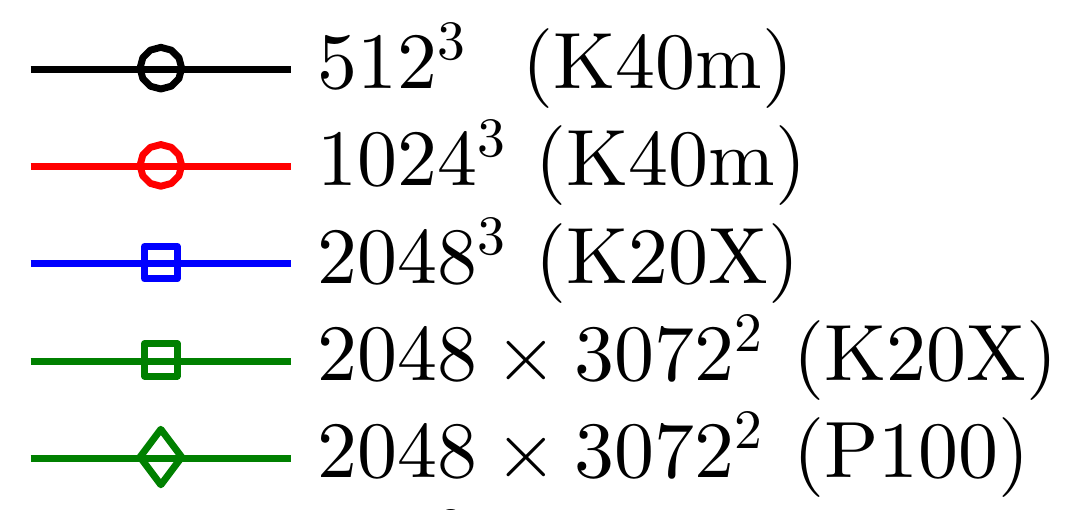}
\caption{Performance of AFiD on CPU and GPU. (a) Speedup. (b) Wall clock time per time-step versus the number of cores. For a fixed grid resolution an increase in this computational time is due to increased communication. With increasing grid resolution the required number of computations increases by more than a factor $N$ due to the pressure solver. (c) CPU time per grid point per time-step for the different test cases. The symbol color indicates the grid size, the solid and open symbols indicate whether the test was performed on CPU or GPU, while the symbol indicates the CPU/GPU model.} \label{fig:scaling2}
\end{centering}
\end{figure}

Given the processor count, the code is able to find the optimal process grid configuration. This is very important in production runs to efficiently use the allocated resources.
The code will factor the total number of MPI tasks and try all the possible configurations, executing the transpose communication routines for a single substep (six transposes required for the Poisson solver), including the halo-exchange time. For the CPU version we measured the performance of the full simulation code to determine the most promising configuration. For the GPU version, with a low processor count the optimal configurations are generally of the form $1\times N$. 

\small
\begin{verbatim}
In auto-tuning mode......
 factors: 1 2 3 6 9 18
 processor grid 1 by 18 time=0.3238s
 processor grid 2 by 9 time=0.8386s
 processor grid 3 by 6 time=0.9210s
 processor grid 6 by 3 time=0.9363s
 processor grid 9 by 2 time=0.8577s
 processor grid 18 by 1 time=0.5901s
 the best processor grid is probably 1 by 18
\end{verbatim}
\normalsize

This is because four of the six transpositions are among processors in the first dimension (xy and xz directions), while only two transpositions are among processors in the second dimension (yz direction). Thus, a $1\times N$ will minimize the amount of data that must be communicated between processors during the Poisson solver transpose routines. However, as the processor count is increased, the halo-exchange time becomes more dominant, and the best configuration becomes the two-dimensional decomposition which minimizes the halo-exchange communications:

\small
\begin{verbatim}
In auto-tuning mode......
factors: 1 2 4 8 16 32 64 128 256 512 1024
 processor grid 2 by 512 time=0.433s
 processor grid 4 by 256 time=8.241E-002s 
 processor grid 8 by 128 time=4.342E-002s 
 processor grid 16 by 64 time=3.173E-002s
 processor grid 32 by 32 time=3.014E-002s
 processor grid 64 by 16 time=4.255E-002s 
 processor grid 128 by 8 time=6.577E-002s 
 processor grid 256 by 4 time=0.121s 
 processor grid 512 by 2 time=0.230s 
 the best processor grid is probably 32 by 32
\end{verbatim}
\normalsize

It is also important to notice that the shape of the decomposition affects the strong scaling since the decomposition changes from 1D to 2D when increasing the processor count.

\subsection{Performance comparison} \label{PerformanceComparison}
The runs were performed on two GPU accelerated systems, the accelerator island of Cartesius at SURFsara and Piz Daint at the Swiss National Supercomputing Centre (CSCS). The accelerator island of Cartesius consists of 66 Bullx B515 GPGPU accelerated nodes, each with two 8-core 2.5 GHz Intel Xeon E5-2450 v2 (Ivy Bridge) CPUs, 96 GB of memory and two 12GB NVidia Tesla K40m GPUs. Every node has a FDR InfiniBand adapter providing 56 Gbit/s inter-node bandwidth.

Piz Daint was originally a Cray XC30 with 5,272 nodes, each with an 8-core Intel Xeon E5-2670 v2 processor, 32 GB of system memory and a 6GB NVidia K20X GPU. 
It has been recently upgraded to a Cray XC50. The compute nodes have now a 12-core Intel Xeon E5-2690 v3 processor, 64 GB of system memory and a 16GB Nvidia P100 GPU. The new Pascal P100 GPU has 720GB/s of peak memory bandwidth (and can sustain more than 500GB/s in the STREAM benchmark) and more than 5 TF of double precision performance. The network is the same before and after the upgrade, and it uses the Aries routing and communications ASICs and a dragonfly network topology. Piz Daint is one of the most efficient petaflop class machines in the world: in the Green 500 list published in November 2013, the machine with XC30 nodes was able to achieve 3186 MFlops/W with level 3 measurements, the most accurate available. In November 2016 with the upgraded XC50 nodes, the machine was able to achieve 7453 MFlops/W, more then doubling the power efficiency. 

We measured the performances of the new accelerated code and compared them to the performance reported in \cite{poe15cf} on the Curie thin nodes (dual 8-core E5-2680 Sandy Bridge EP 2.7GHz with 64GB of memory and a full fat tree Inifiniband QDR network) and with new measurement on Cartesius Haswell thin node islands (2 $\times$ 12-core 2.6 GHz Intel Xeon E5-2690 v3 Haswell nodes) with 64GB of memory per node and 56 Gbit/s inter-node FDR InfiniBand, with an inter-island latency of 3$\mu$s.

Figure \ref{fig:scaling2} shows the scaling data obtained for the CPU and GPU version of the code. The figure shows that both the CPU and GPU version of the code show strong scaling on grid ranging from $512^3$ up to $2048\times3072\times3072$. The required number of GPUs to obtain the same wall-clock time than with the CPU version of the code is much smaller, see figure \ref{fig:scaling2}c. The figure also reveals that we now obtain a better performance and scaling with the CPU version of the code than before, see \cite{poe15cf}. If we focus on the $2048^3$ grid (Table \ref{table:scaling}), we can notice how the wall time with 128 $P100$ GPUs are getting performance very close to the CPU code using about $6K$ cores and while the CPU code is reaching a plateau in efficiency, the GPU code can still scale very well and bring the wall clock time to level unreachable by the CPU version. Since wall clock time is a very important metric for DNS this is a crucial benefit of the GPU version of the code. The new $P100$ GPUs on the upgraded Piz Daint almost double the performance of the code compared to the previous $K20x$ GPUs : since the network stayed the same, but the performance of the GPU increased, the strong scaling is slightly worse. 
\begin{table}
\renewcommand{\arraystretch}{1.3}
\setlength{\tabcolsep}{1pt}
\caption{Wall clock time per step on a $2048^3$ grid. In the CPU simulations there are as many MPI tasks as cpu cores. In the GPU simulation, there are as many MPI tasks as GPUs.}
\label{table:scaling}
\centering
\begin{tabular}{|c|c|c|c|c|c|c|c|}
\hline
 &\multicolumn{2}{|c}{ \bfseries Curie}& \multicolumn{2}{|c}{\bfseries Cartesius} & \multicolumn{3}{|c|}{ \bfseries Piz Daint}\\
 &\multicolumn{2}{|c}{ Xeon E5-2680}& \multicolumn{2}{|c|}{Xeon E5-2690} & { Tesla} &{ K20x} & P100\\
\hline
\hline
Nodes&Cores&Time&Cores&Time&GPUs&Time&Time\\
\hline
32&-&-& 768 & 18.70s& -&-&-\\
64&1024&18.10s& 1536 & 9.58s&-&- &-\\
128&2048&11.18s& 3572 & 5.25s& -&-&2.57s\\
256&4096&6.38s& 6144& 2.82s& 256&2.85s&1.42s\\
512&8192&3.33s& 12288 & 2.03s& 512&1.40s&0.85s\\
1024&16384&2.77s& -& -& 1024&0.74s&0.46s\\
2048&-&-& -& -& 2048&0.50s&-\\ 
4096&-&-& -& -& 4096&0.34s&-\\
\hline
\end{tabular}
\end{table}

Table \ref{table:scaling_pizdaint} shows a comparison between the old XC30 nodes and the new XC50 nodes on Piz Daint for a larger problem on 
a $2048\times3072\times3072$ mesh. We can notice the switch from the 1D decomposition to the 2D decomposition when increasing the processor count.
The larger memory on the P100 (16GB) vs K20X (6GB) makes it possible to run this simulation on 180 nodes and will also allow the use of even finer meshes.

\begin{table}
\renewcommand{\arraystretch}{1.3}
\caption{Wallclock time per step on a $2048\times3072\times3072$ grid. Comparison between the XC30 nodes (Tesla K20X GPU)s vs XC50 nodes (Tesla P100 GPUs).} 
\label{table:scaling_pizdaint}
\centering
\begin{tabular}{|r|r|r|r|}
\hline
Nodes&Configuration&K20X&P100\\
\hline
180&$1\times180$&-& 4.25s\\
256&$1\times256$&-& 2.7s\\
512&$1\times512$&-& 1.95s\\
640&$64\times10$&2.4s& -\\
1024&$64\times16$&1.58s& 1.00s\\
2048&$64\times32$&0.88s& 0.5s\\
4096&$64\times64$&0.57s& 0.36s\\
\hline
\end{tabular}
\end{table}

\section{Validation}\label{section_validation}

All the cases that are shown in this section have been performed with both the CPU and GPU version of the code. The GPU results are identical to the CPU results up to machine precision, which indicates that the CPU and GPU versions of the code are consistent.

\begin{figure*}
\centering
\subfigure{\includegraphics[width=1.\textwidth]{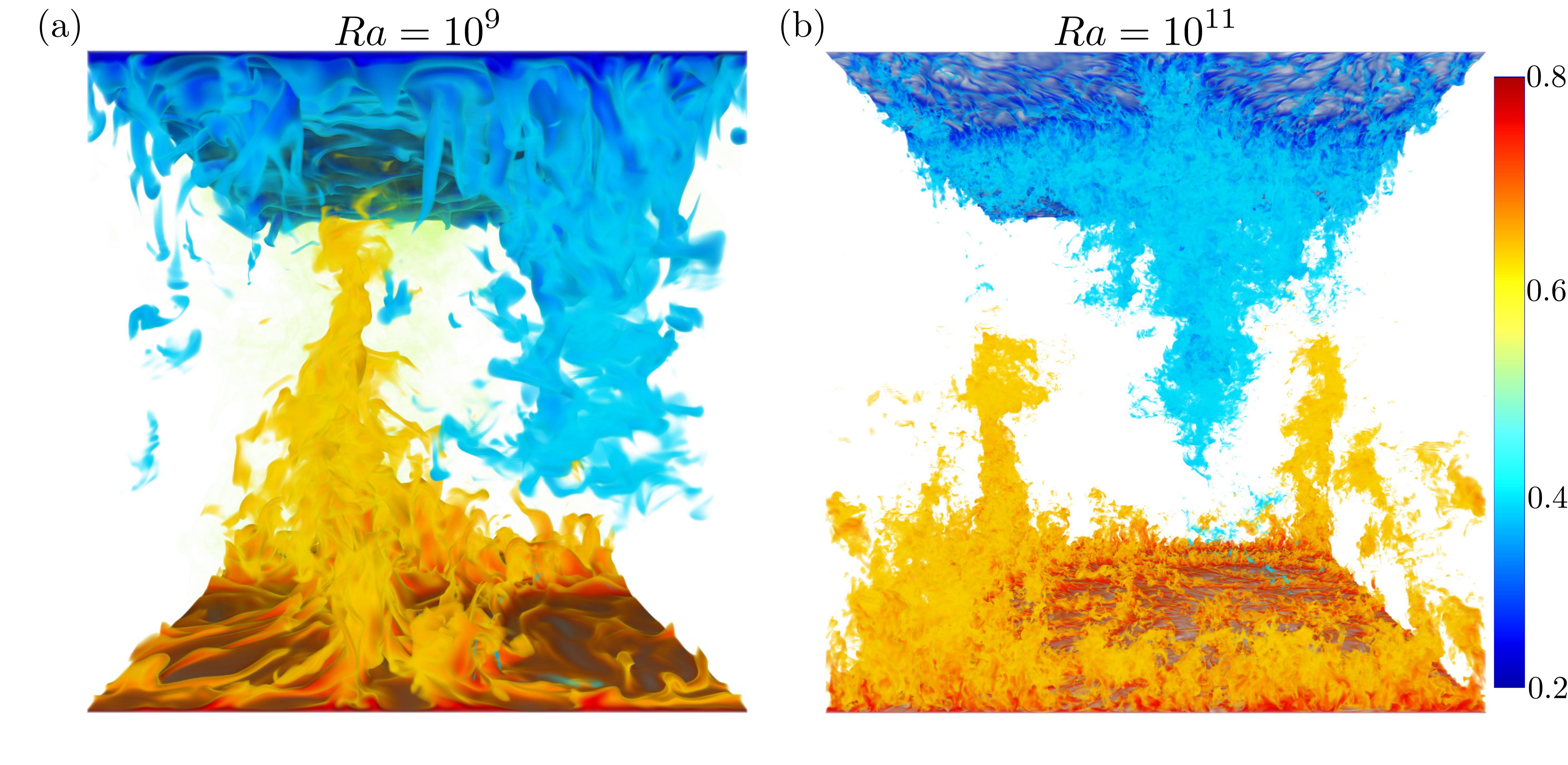}}
\caption{Visualization of the temperature field at $Ra=10^9$ and $Ra=10^{11}$ (rendered on a smaller grid than used during the simulation) for $Pr=1$ in a horizontally periodic $\Gamma=1$ cell. The colorbar indicates the non-dimensional temperature, with the range of 0.2 (blue) to 0.8 (red). }
\label{fig:snapshot}
\end{figure*}

\subsection{Rayleigh-B\'enard convection}

\begin{figure}[!tb]
\centering
\subfigure{\includegraphics[width=0.65\textwidth]{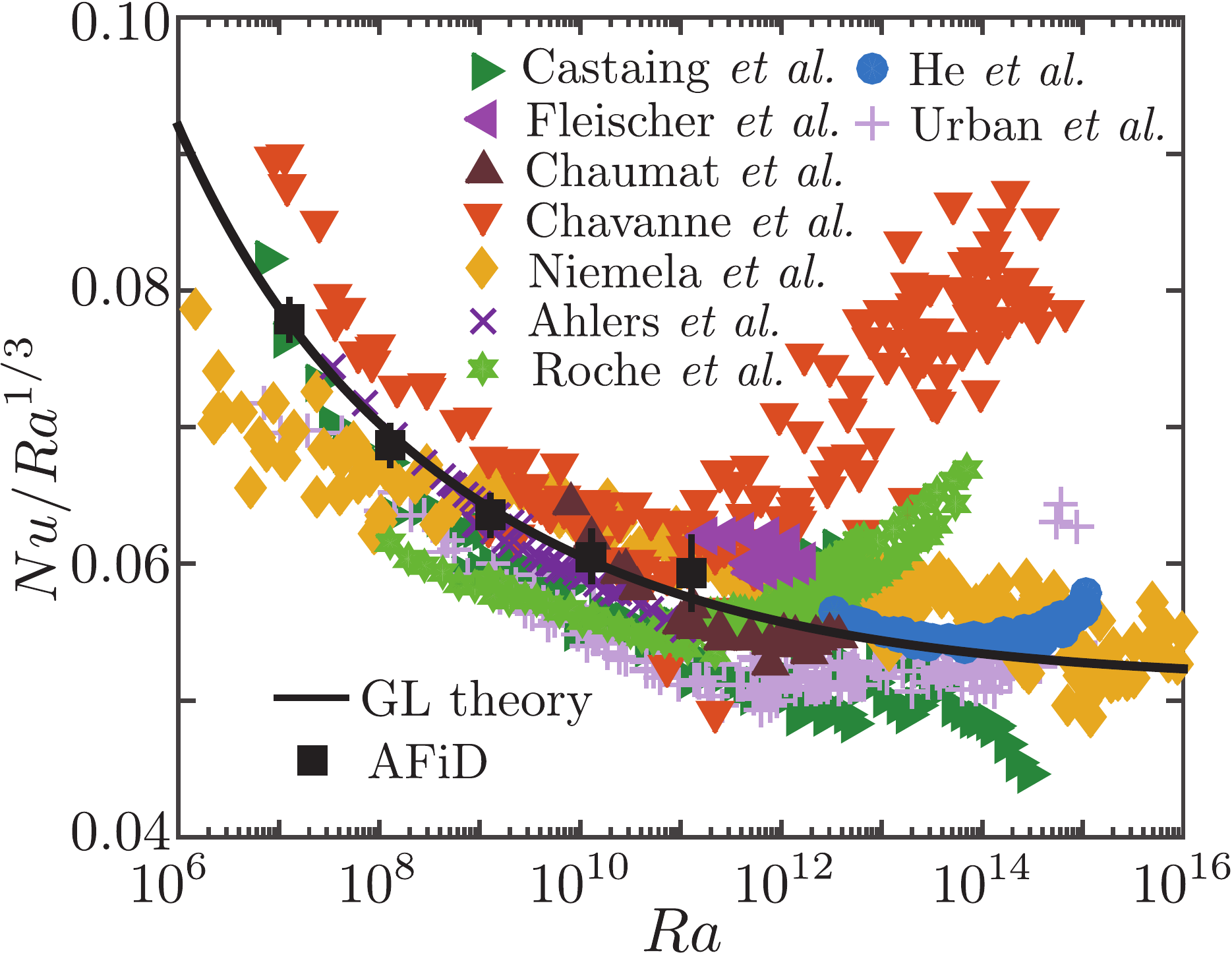}}
\caption{The dimensionless heat flux, i.e. the Nusselt number $Nu$, as a function of the dimensionless temperature difference between the plates, i.e. the Rayleigh number $Ra$, obtained using AFiD, in the compensated way, in comparison with the Grossmann-Lohse theory \cite{gro00,gro01,ste13} and with the experimental data from Castaing \emph{et al.}\cite{cas89}, Roche \emph{et al.} \cite{roc10}, Fleischer \& Goldstein \cite{fle02}, Chaumat \emph{et al.} \cite{cha02}, Chavanne \emph{et al.} \cite{cha01}, Niemela \emph{et al.} \cite{nie00}, Ahlers \emph{et al.} \cite{ahl09c,ahl12b}, He \emph{et al.} \cite{he12}, and Urban \emph{et al.} \cite{urb11,urb12}. The experimental data and GL theory presented in this figure are the same as in Ref.\ \cite{ste13}.}
\label{fig:gl}
\end{figure} 

We simulated Rayleigh-B\'enard convection in an aspect ratio $\Gamma=L/H=1$ cell, where $L$ indicates the streamwise and spanwise domain lengths compared to the domain height $H$. The control parameters of the system are the non-dimensional temperature difference between the plates, i.e.\ the Rayleigh number $Ra$, and the fluid Prandtl number, see Ref.~\cite{ahl09,poe15cf} for more details. 

To test the code we look at the main response parameter of the Rayleigh-B\'enard system, which is the non-dimensional heat transport between the two plates, i.e.\ the Nusselt number. Table \ref{table:rb} shows the simulation details and the extracted Nusselt number for each simulation. In Fig.\ \ref{fig:snapshot} we show snapshots of the flow obtained at different Rayleigh. The figures reveal that the flow structures rapidly decrease with increasing Rayleigh. Fig.\ \ref{fig:snapshot} shows three dimensional visualizations, which show that the GPU code works well and that the turbulent flow structures become much smaller with increasing Rayleigh, illustrating the need of powerful computer codes to simulate very high Rayleigh number flows. 

In Fig.\ \ref{fig:gl}, we show the obtained Nusselt versus Rayleigh compared against experimental data \cite{cas89,roc10,fle02,cha02,cha01,nie00,ahl09c,ahl12b,he12,urb11,urb12} and the predictions by the Grossmann-Lohse theory \cite{gro00,gro01,ste13}. The figure shows that experiments, simulations and theory are in very good agreement with each other up to $Ra=10^{11}$. This figure also shows that there are two facilities (in Grenoble \cite{cha97,roc01,roc10}) which show an increased Nusselt number already around $Ra=5\times10^{11}$, while other experiments (in G\"ottingen \cite{he12,he12a,ahl12b,he15}) show this transitions around $Ra^*_{1}\approx2\times10^{13}$ and $Ra^*_{2}\approx7\times10^{13}$. There is no clear explanation for the mentioned disagreement although it is conjectured that unavoidable variations of the Prandtl number \cite{ahl09,ste11}, finite conductivity \cite{ver04,bro05,ahl09,ste11} of the horizontal plates and sidewall \cite{ahl00,ver02,ste14}, non Oberbeck-Boussinesq effects \cite{ahl06,ahl07,ahl08,sug09,hor13}, i.e.\ the dependence of the fluid properties on the temperature, and even wall roughness \cite{sal14,wag15} and temperature conditions outside the cell might play a role. So far the origin of this discrepancy could never be settled, in spite of major efforts \cite{ahl09,chi12}.

In order to help to clarify these issues it is important to perform DNS with the precise assignment of the temperature boundary conditions (i.e.\ strictly constant temperature horizontal plates and adiabatic sidewall), infinitely smooth surfaces and unconditional validity of the Boussinesq approximation, i.e.\ the fluid properties do not depend on the temperature, which is hard coded in the model equations. In addition, in contrast to experiments, numerical simulations of turbulent flows have the huge advantage that all quantities of the flow are fully accessible while it is possible to adjust the control parameters arbitrarily with the goal to better understand the physics of the system. Our desire to study the transition to the ultimate Rayleigh-B\'enard convection in simulations motivates our development of ever more powerful simulation codes.

\begin{table}
\renewcommand{\arraystretch}{1.3}
\caption{The employed Rayleigh numbers $Ra$ and grid resolution in the horizontal $N_z \times N_y$ and wall-normal  $N_x$ directions, and the extracted $Nu$ from the simulations. For all cases Prandtl numbers $Pr$ and aspect ratio $\Gamma$ are unity.} 
\label{table:rb}
\centering
\begin{tabular}{ccc}
\hline
$Ra$&$N_z \times N_y\times N_x$&$Nu$\\
\hline
$10^7$& $256\times256\times192$&17.17\\
$10^8$& $384\times384\times256$&32.20\\
$10^9$& $512\times512\times384$&64.13\\
$10^{10}$& $768\times768\times512$&132.68\\
$10^{11}$& $768\times768\times1296$&275.33\\
\hline
\end{tabular}
\end{table}

\subsection{Plane Couette flow}

\begin{figure}[!t]
\centering
\subfigure{\includegraphics[width=0.65\textwidth]{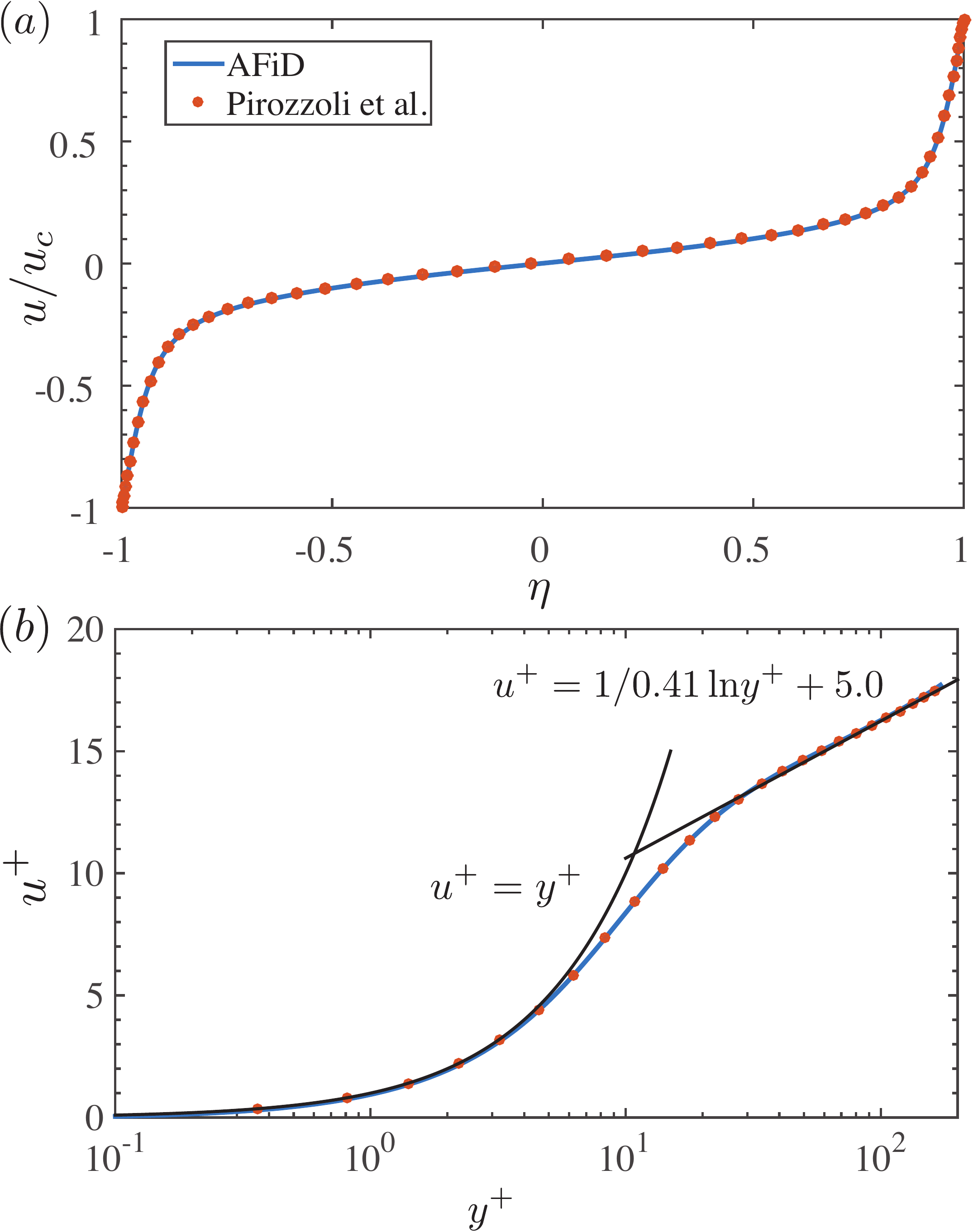}}
\caption{Mean velocity profiles in plane Couette flow scaled with (a) the wall moving velocity $u_c$ and (b) friction velocity $u^+=u/u_\tau$. $\eta$ is the dimensionless wall normal coordinate scaled with the channel half height $h$, in the way that $\eta=0$ corresponds to the channel centreline and $\eta=\pm1$ corresponds to the two walls. $y^+=u_\tau (\eta+1)/\nu$ is the dimensionless distance in wall units. The AFiD results agree excellently with the DNS from Pirozzoli \emph{et al.} \cite{pir14}.}
\label{fig:pc}
\end{figure} 

Now we test the code with the plane Couette configuration at bulk Reynolds number $Re_c=3000$. The two walls here move with the same speed $u_c$ but in the opposite direction. Table \ref{table:pc} shows the employed parameters and the output friction velocity. To capture the large scale structure of plane Couette, rather large domain size $18\pi h \times 8 \pi \times 2h$ has to be implemented, where $h$ is the half height of the channel. The resulting friction Reynolds number $Re_\tau=171$, which is in excellent agreement the previous study under the same confition \cite{pir14}.

\begin{figure*}[!t]
\centering
\subfigure{\includegraphics[width=0.9\textwidth]{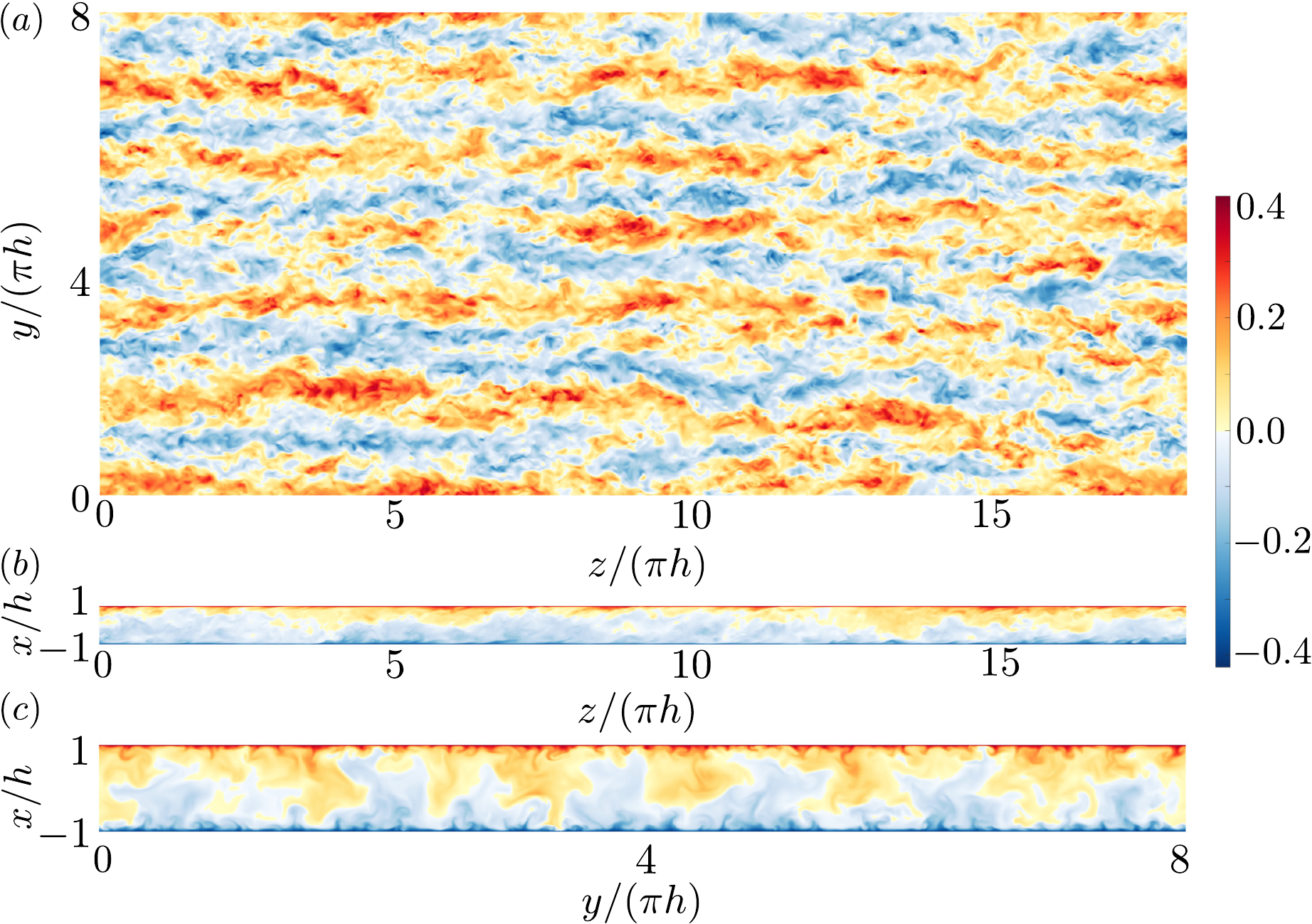}}
\caption{Contours of the instantaneous streamwise flow velocity in (a) the channel centre plane, (b) cross-spanwise view, and (c) cross-streamwise view. }
\label{fig:pcflow}
\end{figure*} 

The streamwise mean velocity profile is shown in figure \ref{fig:pc}, normalized with either the wall velocity $u_c$ or friction velocity $u_\tau$. Again, excellent agreement has been found between the current study and Ref. \cite{pir14}. For figure \ref{fig:pc} (b), two clear layers can be identified. When $y^+<5$, the profile follows $u^+=y^+$, which is called the viscous layer; When $50<y^+<171$, a clear logarithmic layer is seen, with $u^+=1/\kappa\, \mathrm{ln} y^+ +C$, where $\kappa \approx0.41$ and $C\approx5.0$.   

Figure \ref{fig:pcflow} shows the large scale flow structure of the plane Couette flow. Distinctive patterns of high and low speed steaks are evident, which maintains the coherence along the whole streamwise length of the channel, while also showing some meandering. The spanwise width of the large scale structure is of (4-5)h, as also shown in previous studies \cite{avs14,pir14}. The above findings demonstrate the necessity for extreme large box to capture the biggest structure which might be present in the plane Couette flow, and this is reason that biggest DNS done so far in the plane Couette flow is only at $Re_\tau \approx 1000$ \cite{pir14}, while for channel flow it is at $Re_\tau \approx 5200$ \cite{lee15}.

With this GPU version of the code, our goal is to study on the one hand even bigger box size, which will help understand how big the large scale structure can really be. On the other hand, we want to study higher Reynolds number plane Couette flow, which will help understand how far the logarithmic layer can extend and whether attached eddy hypothesis \cite{mar10b} can work for plane Couette flow.

\begin{table}
\renewcommand{\arraystretch}{1.3}
\caption{List of parameters for the plane Couette flow case. Here $Re_c=hu_c/\nu$ is the bulk Reynolds number and $h$ is the channel half height, $u_c$ the moving velocity of wall, $\nu$ the kinematic viscosity. The second column shows the computational box. The third column shows the grid resolution. The last column is the friction Reynolds number $Re_\tau=hu_\tau/\nu$, where $u_\tau$ is defined as $u_\tau=\sqrt {\tau_w/\rho}$, in which $\tau_w$ is the wall shear stress. Note that $Re_\tau$ is output of the system.} 
\label{table:pc}
\centering
\begin{tabular}{cccc}
\hline
$Re_c$&$L_z\times L_y \times L_x$&$N_z \times N_y\times N_x$&$Re_\tau$\\
\hline
$3000$&$18\pi h \times 8\pi \times 2 h $ & $1280\times1024\times256$ &171 \\
\hline
\end{tabular}
\end{table}

\section{Conclusions and future plans} \label{section_conclusions}
In this paper we presented a GPU accelerated solver that can be used to study various wall-bounded flows \cite{ahl09,loh10,far14,gro16,mar10b,smi11,jim12,smi13,mar13}. Our work is motived by the need to simulate more extreme turbulent flows and inspired by the observation that while high performance computing shifts towards GPUs and accelerators, obtaining an efficient GPU code, that is faster than is CPU, is thought to be a very time consuming, code specific, undertaking. In this paper we showed that to port CPU code to the GPU, only ``minimal effort'' is required and one can achieve an order of magnitude improvement in wall clock time when comparing the GPU code to the CPU one. In this work we presented some efficient coding techniques, such as overloaded sourced allocation and how module use/renaming can be used to avoid modifying loop contents that have not been covered elsewhere. In addition, we point out that this approach allows for easy code validation, since every subroutine can be examined to produce the same results as the original CPU code up to the machine precision. This approach is generally applicable and an eye opener for many scientists thinking about GPU porting. 

The need to obtain ever more efficient codes is illustrated by example use cases of high Rayleigh number turbulent Rayleigh-B\'enard convection and high Reynolds number plane Couette flow. We have shown in \textsection \ref{section_validation} that our code can work perfectly with both of the flows. Our motivation to perform higher Reynolds and high Rayleigh number simulations are much more demanding than the tested cases. From this point of view, the GPU code developed here is highly required.

Previous work to parallelize second-order finite-difference solvers allowed us to reach extremely high Reynolds numbers in Taylor-Couette flow \cite{ost16jfm}, and also to simulate Taylor-Couette flow with riblets in the flow direction and notches in perpendicular to the flow direction to disentangle the effects of roughness on the torque \cite{zhu16jfm,zhu16b}. For Rayleigh-B\'enard convection we have used the AFiD code to simulate unprecendetly large horizontal domains to investigate the formation of   Rayleigh-B\'enard superstructures \cite{ste17}. With the GPU code described here, the capability of the code is improved ever further. Initial works have been started to simulate the Rayleigh-B\'enard flows with external shearing by using the GPU version. It should be pointed out that the code can also be used to simulate other wall-turbulent flow at high Reynolds number such as channel flow.

In order to further expand the capabilities of the code, we are going to work on several fronts. The first one will be to utilize the CPU cores, which are completely idle in the basic code version described here, together with the GPUs in the implicit part of the solver. We are more interested in the CPU memory than the CPU flops, but depending on the node configuration, the CPU cores can give a good performance boost. Subdividing each vertical domain in two subdomains, we can process one on the CPU and one on the GPU. The relative size of the subdomains can be determined at runtime, since the workload per cell is constant. The split is in the outermost dimension (z) and requires additional halo exchanges (but these are local memory transfers of contiguous data between GPU and CPU, so no network is involved). A preliminary version of this hybrid CPU-GPU code is available in the open-source code, which will be discussed in detail in a forthcoming paper, and is currently being tested.

Since writing the HDF5 files to disk could be time consuming, we are thinking about making this process asynchronous, once the solution is copied to CPU memory, the GPU can advance the solution while the CPUs complete the I/O.
 
Another way to optimize the simulations is to use a multiple resolution approach, using a grid for the temperature field with a higher spatial resolution than that for the momentum, as integrating both fields on a single grid tailored to the most demanding variable produces an unnecessary computational overhead. This approach gives significant savings in computational time and memory occupancy as most resources are spent on solving the momentum equations (about $80\%-90\%$). To assure stable time integration of the temperature field we use a separate refined time-step procedure for the temperature field. The full details of the strategy are described in \cite{ost15}.

The GPU code will be released as open source and it will be available for download at {\color{blue}\burl{http://www.afid.eu}}.

\section{Acknowledgments}
This work was supported by a grant from the Swiss National Supercomputing Centre (CSCS) under project ID g33 and by the Netherlands Center for Multiscale Catalytic Energy Conversion (MCEC), an NWO Gravitation program funded by the Ministry of Education, Culture and Science of the government of the Netherlands, FOM (Foundation for Fundamental Matter) (FOM) in the Netherlands, and the ERC Advanced Grant ``Physics of boiling''. We also acknowledge PRACE for awarding us access to FERMI and Marconi based in Italy at CINECA under PRACE project number 2015133124 and 2016143351 and NWO for granting us computational time on Cartesius cluster from the Dutch Supercomputing Consortium SURFsara and for the continuous support we get from SURFsara on code development.

\bibliography{literatur}
\bibliographystyle{elsarticle-num}

\end{document}